\documentclass[aps,prl,twocolumn,showpacs,preprintnumbers,amsmath,amssymb,superscriptaddress]
{revtex4}
\usepackage{graphicx}
\usepackage{dcolumn}
\usepackage{bm}

\begin{document}
\title
{Intrinsic spin Hall effect in platinum metal}
 
\author{G. Y. Guo}
\affiliation{Department of Physics and Center for Theoretical Sciences, National Taiwan University, Taipei 106, Taiwan}
\author{S. Murakami}
\affiliation{Department of Physics, Tokyo Institute of Technology, 
2-12-1 Ookayama, Meguro-ku, Tokyo 152-8551, Japan}
\author{T.-W. Chen}
\affiliation{Department of Physics and Center for Theoretical Sciences, National Taiwan University, Taipei 106, Taiwan}
\author{N. Nagaosa}
\affiliation{CREST, 
Department of Applied Physics, University of Tokyo, Tokyo 113-8656, Japan}
\affiliation{Correlated Electron Research Center, National
Institute of Advanced Industrial Science and Technology, 1-1-1,
Higashi, Tsukuba, Ibaraki 305-8562, Japan}

\date{\today}   

\begin{abstract}
Spin Hall effect (SHE) is studied with
first-principles relativistic band calculations for 
platinum, which is one of the most important materials for metallic
SHE and spintronics. 
We find that intrinsic spin Hall conductivity  (SHC)
is as large as $\sim 2000 (\hbar/e)(\Omega {\rm cm})^{-1}$
at low temperature, and decreases down to 
$\sim 200 (\hbar/e)(\Omega {\rm cm})^{-1}$ at room temperature.
It is due to the resonant contribution from the spin-orbit splitting
of the doubly degenerated $d$-bands at high-symmetry
$L$ and $X$ points near the Fermi level. By modeling these near-degeneracies 
by an 
effective Hamiltonian, we show that SHC has a peak near the Fermi energy and
that the vertex correction due to impurity scattering vanishes. 
We therefore argue that the large SHE observed
experimentally in platinum is of intrinsic nature.
\end{abstract}
\maketitle

Spin Hall effect (SHE), i.e., the transverse 
spin current generation by the 
electric field, is an issue of intensive current interests both 
theoretically and experimentally since the theoretical proposal for 
its intrinsic mechanism in semiconductors \cite{mur03,sin04}. 
This effect enables us to control spins without
magnetic field or magnetic materials, which is a crucial step for 
spintronics. 
In addition to semiconductors, the SHE in {\it metallic} systems
is currently attracting interest, stimulated by 
experiments on 
the SHE or inverse spin Hall effect (ISHE),
i.e. the transverse voltage drop due to the spin current 
\cite{Saitoh06,Kimura06,Valenzuela06}.
SHE/ISHE in metals has the following 
importance and advantages compared with that in semiconductors; 
(i) A contact with a ferromagnetic metal
does not suffer from conductance mismatch \cite{Schmidt00}, 
and one can make use of the spin-polarized current supplied from it. 
Thus, 
techniques developed in metallic spintronics can be 
utilized. (ii) The spin Hall conductivity (SHC) 
is much larger than that in semiconductors. 
The value of SHC obtained in Ref.~\cite{Kimura06} is 4 orders of 
magnitude larger than that in GaAs \cite{kat04}. 
Naively this appears to be attributed to the large number of carriers, 
whereas the band structure is important as we discuss below. (iii) 
The Fermi degeneracy temperature is much higher than room temperature, 
and hence quantum coherence is more robust against thermal 
agitations than in semiconductors.
We note that the spin diffusion length is relatively small in 
metals, e.g. 10 nm in platinum (Pt) \cite{Kimura06}, causing 
fast decay of the SHE signal. 
However, it is not a crucial obstacle for observation 
and application, by designing the device       
as demonstrated in Ref.~\cite{Kimura06}.

Compared with the recent experimental advances in metallic SHE, 
its theoretical understanding is still lacking and is urgent. 
Among metallic systems, Pt shows remarkably large SHE surviving 
even up to room temperature \cite{Saitoh06,Kimura06},
whereas aluminum and copper show relatively tiny SHE \cite{Valenzuela06}. 
The SHC in Pt at room temperature is 240 $(\hbar/e)(\Omega {\rm cm})^{-1}$, 
ten times larger than that of aluminum at 4.2K. 
In ~\cite{Kimura06} this difference is attributed to a
magnitude of spin-orbit coupling for each metal. 
However, Pt 
seems to be special even among heavy elements, and the SHC does not
simply scale with the size of the spin-orbit coupling.
Such behavior cannot be explained within the extrinsic mechanism
\cite{dya71,eng05,shc05},
where material properties are represented by 
a few parameters such as the size of the spin-orbit coupling.
This material dependence strongly suggests a crucial role of intrinsic 
contributions, which has been largely overlooked. 
It is thus highly desired to study the intrinsic SHE of Pt as a representative 
material for metallic SHE. This analysis 
opens up the possibility to theoretically design the 
SHE in metallic systems.

This discussion on separating intrinsic and extrinsic mechanisms
is analogous to the long-standing debates
on the anomalous Hall effect (AHE)\cite{AHE,AHE2,smit,berger}. 
In semiconductors, there have been experimental reports on the SHE
in $n$-GaAs \cite{kat04}, $p$-GaAs \cite{wun04} and
$n$-type InGaN/GaN superlattices \cite{cha07}.  
It is now recognized that the 
SHE in $n$-type GaAs is due to the extrinsic mechanisms, i.e., skew 
scattering and side-jump contributions \cite{ino04,eng05}, 
while that in $p$-type GaAs is mostly intrinsic \cite{mur04b,Onoda05}. 
In metals, the conventional understanding has been that the skew 
scattering is dominant in AHE. However, recent studies have revealed 
that the intrinsic contribution can be dominant 
for AHE in metals when the  $\sigma_{xy}$ is of the order of 10$^3$ 
$(\Omega{\rm cm})^{-1}$
and the conductivity $\sigma_{xx}$ is in the range of $ \sim 10^4-10^6
\Omega^{-1}$
cm$^{-1}$ \cite{Onoda06}. This dominant contribution of intrinsic mechanism
is confirmed by the detailed comparisons 
between the first-principles calculations \cite{Fang03,Yao04,yao07} 
and experiments
\cite{asamitsu}. 

In this Letter we present an {\it ab initio} calculation
for the SHC in Pt, and its analysis based on an effective Hamiltonian. 
We find that there are near-degeneracies near the Fermi level ($E_F$) at 
high-symmetry $X$ and $L$ points in the 
Brillouin zone (BZ) for the fcc lattice. 
They give a prominent enhancement of 
SHC in Pt. We determine an 
effective Hamiltonian near $X$ and $L$ points,
and demonstrate robustness of the SHE against impurities.

The band structure of Pt is calculated using a 
fully relativistic extension~\cite{ebe88}
of the all-electron linear muffin-tin orbital
method~\cite{and75} based on the density functional theory with 
local density approximation~\cite{vos80}. 
The lattice constants for Pt and Al used 
are 3.92 and 4.05 \AA, respectively.
The basis functions used are $s$, $p$, $d$ and $f$ 
muffin-tin orbitals for Pt but $s$,
$p$, and $d$ muffin-tin orbitals for Al~\cite{and75}.
In the self-consistent band structure calculations, 
89 $k$-points in the fcc irreducible wedge (IW)
of the BZ were used in the 
BZ integration. 
The SHC is evaluated by 
the Kubo formula \cite{guo05}.
A fine mesh of  60288 $k$-points on a larger IW 
(three times the fcc IW) is used.  These correspond to the  
division of the $\Gamma X$ line into 60 segments.
Comparison with test calculations with 102315 $k$-points 
(72 divisions of the $\Gamma X$ line) for Pt indicate that 
the calculated SHC converges within 1 \%. 

\begin{figure}[h]
\includegraphics[width=8cm]{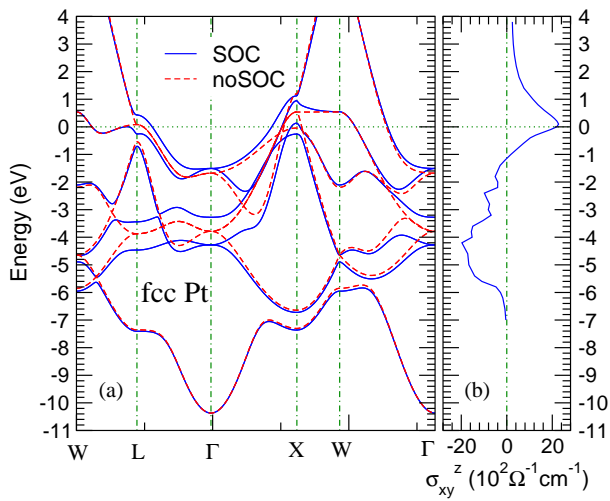}
\caption{\label{bs} (color online) (a) 
 Relativistic band structure and (b) spin Hall conductivity 
 of fcc Pt. The zero energy and the dotted line
 is the Fermi level.
 The dashed curves in (a) are the scalar-relativistic
band structure.}
\end{figure}

Fig. 1 shows the relativistic band structure of Pt, and also the 
SHC ($\sigma_{xy}$) as a function of $E_F$. 
Remarkably, the SHC peaks at the true Fermi level (0 eV), with a large 
value of
2200 $(\hbar /e)(\Omega{\rm cm})^{-1}$. This gigantic value of
the SHC is orders of magnitude larger
than the corresponding value in $p$-type semiconductors 
Si, Ge, GaAs and AlAs~\cite{guo05,yao05}. Furthermore, the calculated 
SHC in simple metal Al is only $-$17 $(\hbar /e)(\Omega{\rm cm})^{-1}$, 
being two orders of magnitude smaller than that of Pt.
Interestingly, the SHC in Pt decreases 
monotonically as the $E_F$ is artificially raised 
and becomes rather small above 3.0 eV. When
the $E_F$ is artificially lowered, the SHC
also decreases considerably, and changes its
sign at $-$1.1 eV. As the $E_F$ is further lowered,
the SHC increases in magnitude again,
and becomes peaked at $-$4.2 eV with a large value of
$-$1970 $(\hbar /e)(\Omega{\rm cm})^{-1}$. The SHC
decreases again when the $E_F$ is further lowered,
and finally becomes very small below $-$6.0 eV.
Note that the bands below $-$8.0 eV and also above 2.0 eV
are predominantly of 5$s$ character and the effect of the
spin-orbit coupling is negligible. 

We notice that a peak in the SHC appears at the double
degeneracies on the $L$ and $X$ points near 
$E_F$ (0 eV) in the scalar-relativistic band structure 
(i.e., without the spin-orbit coupling) 
while the other peak at $-$4.2 eV occurs near the double
degeneracies at the $L$ and $\Gamma$ points (see Fig.~\ref{bs}).
The double degeneracy (bands 5 and 6) at
$L$ is made mostly (93 \%) of $d_{x'z'}$ and $d_{y'z'}$ ($z'$: threefold 
axis),
being consistent with the point group $D_{3d}$ at $L$. 
The double degeneracy (bands 4 and 5) at
$X$ consists mainly of $d_{x'z'}$ and $d_{y'z'}$ ($z'$: fourfold 
axis),
being consistent with the point group $D_{4h}$.
These double degeneracies are lifted by
the spin-orbit coupling, with a large spin-orbit
splittings ($\sim$ 0.66, 0.93 eV, respectively).

One may attribute the large SHC in Pt to these double 
degeneracies. To see this, let us consider the $k$-resolved
contribution to the SHC, i.e., Berry 
curvature $\Omega_n^z({\bf k})$;
\begin{eqnarray}
\sigma_{xy}^{z} = \frac{e}{\hbar}\sum_{\bf k}\Omega^z({\bf k}) 
= \frac{e}{\hbar}\sum_{\bf k}\sum_n f_{{\bf k}n}\Omega_n^z({\bf k}),\nonumber \\
\Omega_n^z({\bf k}) = \sum_{n'\neq n}
\frac{2{\rm Im}[\langle{\bf k}n|j_x^z|{\bf k}n'\rangle\langle{\bf k}n'|v_y|{\bf k}n\rangle]}
 {(\epsilon_{{\bf k}n}-\epsilon_{{\bf k}n'})^2},
\end{eqnarray}
where the spin current operator $j_x^z = \frac{1}{2}\{s_z, {\bf v}\}$, 
with spin $s_z$ given by
$s_z=\frac{\hbar}{2}\beta\Sigma_z$ ($\beta$,  
$\Sigma_z$: $4\times 4$ Dirac matrices) 
\cite{guo05}.
$f_{{\bf k}n}$ is the Fermi distribution function for the $n$-th band at 
${\bf k}$. ${\Omega_n}^{z}$ is 
an analogue of the Berry curvature for the $n$-th band, and it is
enhanced when other bands come close in energy (i.e.\ near-degeneracy). 
Fig. 2(a) shows clearly that $\Omega^z({\bf k})$ is large only near 
the $L$ and $X$ points. Interestingly, Berry curvature
$\Omega_n^z({\bf k})$ for the doublet bands 4 and 5 near the $X$ point
are large but have opposite signs (Fig.~2(b)). However, because band 
5 near
the $X$ point is unoccupied, only $\Omega_n^z({\bf k})$
for band 4 contributes to the SHC, resulting in the large positive
peak in $\Omega^z({\bf k})$ near the $X$ point (Fig.~2(a)).
Fig.~2(c) shows that the SHC decreases monotonically as the temperature
($T$) is raised. This rather strong temperature dependence is also 
due to the near-degeneracies since the small energy scale is relevant
to the SHC there. Nevertheless, the SHC
$\sigma_{xy}= 240 (\hbar /e)(\Omega{\rm cm})^{-1}$  at $T=300$K is
still large, and is close to the measured value (240)
\cite{Kimura06}.
The SHC for Al at 4 and 300 K is $-17$ 
and $-6 (\hbar /e)(\Omega{\rm cm})^{-1}$,
respectively. The former value is similar to the experimental 
values ($-$27, $-$34) at 4.2 K \cite{Valenzuela06}.

\begin{figure}[h]
\includegraphics[width=7cm]{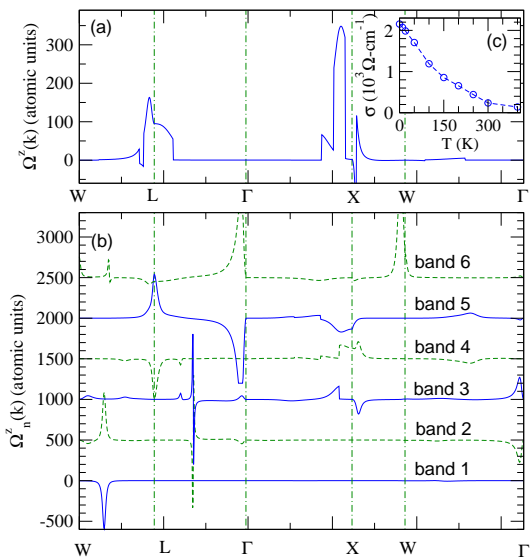}
\caption{\label{k-resolved} (color online) 
(a) Berry curvature $\Omega^z({\bf k})$ at zero temperature, 
and (b) band($n$)-decomposed Berry curvature
$\Omega_n^z({\bf k})$ along the symmetry lines in the fcc Brillouin zone.
In (b), $\Omega_n^z({\bf k})$ for 
the $n$th band has been shifted upwards by $(n-1)\times500$ for clarity.
The inset (c) shows the temperature-dependence of the
spin Hall conductivity $\sigma^{z}_{xy}$.}
\end{figure}

In order to study the role of near-degeneracies in more detail, 
we construct two effective Hamiltonians $H({\bf k})$
for the two doubly degenerate bands at $X$ and $L$ points, 
respectively.
At the $X$ point, by imposing 
the $D_{4h}$ symmetry and the time-reversal symmetry,
the effective Hamiltonian 
with basis $|(x'\mp iy')z'\uparrow\rangle$ and $|(x'\pm
iy')z'\downarrow\rangle$ ($z'$: fourfold axis) can be written 
in terms of 4$\times$4 Clifford $\Gamma$-matrices~($\Gamma^1=\tau_x$,
$\Gamma^2=\sigma_z\tau_y$, $\Gamma^3=\sigma_x\tau_y$,
$\Gamma^4=\sigma_y\tau_y$, $\Gamma^5=\tau_z$) as
$H(\mathbf{k})=\epsilon(\mathbf{k})+\sum^{5}_{a=1}d_a
(\mathbf{k})\Gamma^{a}$.
By expanding the coefficients $d_a$ with respect to the wavenumbers 
$\mathbf{k}'$
measured from $X$ and $L$ points ($\mathbf{k}'=\mathbf{k}-\mathbf{k}_{i}$, 
$i=L,X$), we have constructed the effective 
Hamiltonian. Fitting with the calculated energy bands and wavefunctions, we 
determined the expansion coefficients to $k'^4$ order. 
This effective model is an even function of $\mathbf{k}'$, and 
is similar to the Luttinger model, representing
the valence bands of cubic semiconductors \cite{mur04}, or the valence 
and conduction bands of zero-gap cubic semiconductors \cite{mur04a} near 
the $\Gamma$-point. 
The previous analysis for the $p$-type semiconductors \cite{mur04}
are equally applied.
The effective Hamiltonian has the eigenvalues 
$E_{{\rm l}}({\bf k})=\epsilon({\bf k})-d({\bf k})$, 
and $E_{{\rm u}}({\bf k})=\epsilon({\bf k})+d({\bf k})$ for the 
lower and upper bands, respectively, where 
$d=\sqrt{\sum_{i=1}^{5}d_i^2}$, and these bands
correspond to the heavy-hole and light-hole bands, respectively.
 From Eq.~(35) of
 Ref.~\cite{mur04},
the response of a generalized spin current (corresponding to
$\Gamma^{ab}$) is given by 
\begin{equation}
\sigma_{ij}^{ab} =4\int \frac{d{\bf k}}{(2\pi)^{3}}
(f_{\mathbf{k}{\rm l}}-f_{\mathbf{k}{\rm u}})
G^{ab}_{ij}, 
\label{sigmaijab}
\end{equation}
where $f_{{\bf k}{\rm u}}$ and
$f_{{\bf k}{\rm l}}$ are
the Fermi functions of the upper and the lower bands, and
$G^{ab}_{ij}=\frac{1}{4d^{3}}
\epsilon_{abcde}d_{c}\frac{\partial d_{d}}{\partial k_{i}
}\frac{\partial d_{e}}{\partial k_{j}}
$
where $\epsilon_{abcde}$ is the totally antisymmetric tensor 
with $\epsilon_{12345}=1$.
We flipped the sign of $\sigma_{ij}^{ab}$ because the sign of the 
charge of the carriers is opposite from Ref.~\cite{mur04}.
$G^{ab}_{ij}$ describes the mapping of an area form from
the three-dimensional  $\mathbf{k}$ space to the
five-dimensional $\mathbf{d}$ space. 
It can be regarded as a ``solid angle'' enclosed by the 
${\bf d}$ vector when the wavenumber ${\bf k}$ runs over the domain
between the two Fermi surfaces. Hence, it becomes larger for smaller 
$d({\bf k})=\frac{1}{2}(E_{{\rm u}}-E_{{\rm l}})$. 
The spin operators are given by 
$s^x=\Gamma^{35}/2$, 
$s^y=\Gamma^{45}/2$, and 
$s^z=\Gamma^{34}/2$, 
where $\Gamma^{ab}=\frac{1}{2i}[\Gamma^{a},\ \Gamma^{b}]$.
Using these relations, one can calculate the SHC $\sigma_{xy}^{z}$ from 
Eq.~(\ref{sigmaijab}) by summing over the three $X$ points and four $L$ points.

\begin{figure}[h]
\includegraphics[width=8.0cm]{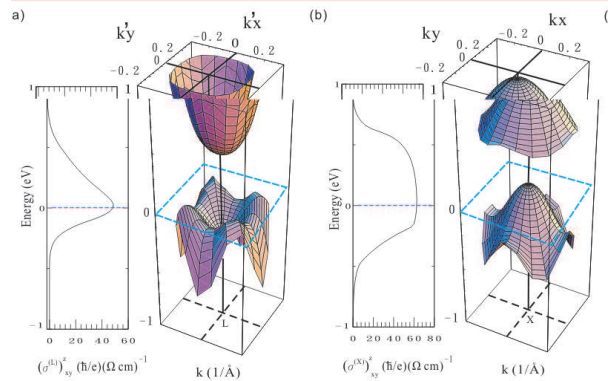}
\caption{\label{SHC-effective} (color online) Spin Hall conductivity of platinum 
calculated from the effective Hamiltonian for (a) the $L$ points
and (b) the $X$ points, as a function of $E_F$ \cite{remark}. 
}
\end{figure}

The next issue is whether 
the contributions from various bands cancel or not. {}From 
Eq.~(\ref{sigmaijab}), the SHC from the 
$X$-points and that from the $L$-points are calculated as 
a function of the $E_F$, as shown in 
Fig.~\ref{SHC-effective}. Here we put a cutoff for the
${\bf k}$-integral as $\pi/(5a)$. 
The integrand is dominated by the contribution 
near the $L$ or the $X$ points, and cancellation does not occur
when the Fermi energy is in the gap.
It is analogous to the zero-gap semiconductors rather than 
GaAs \cite{mur04a,remark}.
Thus we can identify the peaks at $E_F\sim 0$ with the peak of 
the SHC in Fig.~\ref{bs}, and the enhancement of SHC in 
Pt is attibuted to the near-degeneracies at the $L$ and $X$ points.
As is similar to the $p$-type semiconductors \cite{mur04b}, this 
intrinsic SHE
is robust against impurity scattering \cite{note-skew}. 
To see this, we consider dilutely distributed 
short-ranged impurities $V(\mathbf{r})=\sum_{i}V\delta(\mathbf{r}-
\mathbf{r}_{i})$. It is justified in Pt, 
because screening is prominent compared with semiconductors.
Then the vertex corrections from the impurity scattering for the SHC 
vanishes in the clean limit from the following reason. 
Because the effective Hamiltonian satisfies
$H(\mathbf{k})=H(-\mathbf{k})$,
the Green function is an 
even function 
and the current operator is an odd function 
of  $\mathbf{k}'$. 
Then in calculating the SHC from a correlation function 
between the current $j_y$ and the spin current $j_x^z$,
the ladder diagrams from 
impurities cancel between the internal wavenumbers 
$\mathbf{k}'$ and 
$-\mathbf{k}'$ for the current vertex $\mathbf{j}(\mathbf{k})$.
Thus for short-ranged impurities, the SHC in the clean limit is given 
by the intrinsic value from the bare diagram without impurity 
scattering. This justifies our first-principle result even in the
disordered case.
Although it may sound trivial, it is not in general; 
in the Rashba model the vertex correction from impurities 
is relevant, and kills the intrinsic SHC even in the clean limit\cite{ino04}. 

We note that 
$H(\mathbf{k}')=H(-\mathbf{k}')$
 results because we restrict ourselves 
to the even-parity (i.e.\ $d$) orbitals. Thus even when 
we include the higher-order terms in $\mathbf{k}'$ it holds true, and 
the vertex correction vanishes for short-ranged impurities.
When ${\bf k}$ is away from such high-symmetry points, the orbitals 
with odd and even parities are hybridized, and 
the SHC will be cancelled to some extent by the vertex 
corrections by impurities.
Thus for inversion-symmetric systems such as Pt,
it is safe to restrict ourselves to the high-symmetry points.

Discussion on the relevance of the present result to the 
experiment on SHE in Pt \cite{Kimura06} is in order. 
At room temperature 
the magnitude of $\sigma^z_{xy} 
\sim 240$ $(\Omega$cm$)^{-1}$ with the conductivity 
$\sigma_{xx}\sim10^5$ $\Omega^{-1}$cm$^{-1}$, 
corresponds to the ``intrinsic" region in the criterion 
of Ref.~\cite{Onoda06}. This is consistent with the idea 
of ``resonant" Hall effect since the enhanced contribution from the 
near degeneracies at X- and L-points has
been confirmed by the present first-principles calculation. 
Hence it is most probable that the SHE in Pt at room temperature
is due to the 
intrinsic mechanism calculated in this Letter.
On the other hand, at the lowest temperature the system enters
the superclean extrinsic region \cite{Onoda06}, with $\sigma_{xx}$ rising
up to $\sigma_{xx}\sim 10^9$ $\Omega^{-1}$cm$^{-1}$. 
Hence at lowest temperature the skew scattering becomes very large, 
and the SHC cannot be explained only by the intrinsic mechanism.

The authors thank National Science Council and NCTS of ROC for support, 
and also NCHC of ROC for CPU time.
The work was partly supported by Grant-in-Aids 
(Grant Nos. 15104006, 16076205, 17105002, 19740177, and 19019004)
and NAREGI Nanoscience Project from the Ministry of 
Education, Culture, Sports, Science, and Technology of Japan.

{\it Note added}: After submission of the paper, temperature
dependence of the SHC in Pt was measured to be almost constant from $T=300$K 
to $T=0$K~\cite{Vila07}. 
Though it may look different from our scenario, it is consistent 
with it. In ~\cite{Vila07}, the conductivity at 
$T=0$K is $\sigma_{xx}\sim 10^5(\Omega{\rm cm})^{-1}$, much lower than the 
above-mentioned value. This conductivity corresponds to the
self-energy of the order of 10meV,  which is comparable to
room temperature. This implies that the self-energy 
gives a cutoff to the expression of the SHC, and the SHC
remains constant below room temperature.


\begin{thebibliography}{99}
\bibitem{remark} As seen in Fig.1, there is no real gap near $L$ and 
$X$ points while Fig. 3 shows the finite gaps. This is because we take
a rather small value of the cut-off because the fitting by the 
second order terms in $k$ gets worse for 
larger cut-off.  It is also the reason why the obtained value
of SHC for the effective model is one order of magnitude smaller
than the band calculation.
\bibitem{note-skew}
Here we are considering an intrinsic 
region 10$^4$ $(\Omega{\rm cm})^{-1}< \sigma_{xx}< 10^6$ 
$(\Omega{\rm cm})^{-1}$, and skew-scattering 
contribution is neglected.
\end{thebibliography}
\end{document}